# Vortex Laser at Exceptional Point

Xing-Yuan Wang[1], Hua-Zhou Chen[1], Ying Li[1], Bo Li[1], Ren-Min Ma[1,2]

[1] State Key Lab for Mesoscopic Physics and School of Physics, Peking University, Beijing 100871, China.

[2] Collaborative Innovation Center of Quantum Matter, Beijing, China

**Abstract**

The optical vortices carrying orbital angular momentum (OAM) are commonly generated by modulating the available conventional light beam. This article shows that a micro-laser operates at the exceptional point (EP) of the non-Hermitian quantum system can directly emit vortex laser with well-defined OAM at will. Two gratings (the refractive index modulation and along azimuthal direction and the grating protruding from the micro-ring cavity) modulate the eigenmode of a micro-ring cavity to be a vortex laser mode. The phase-matching condition ensures that we can tune the OAM of the vortex beam to be arbitrary orders by changing the grating protruding from the micro-ring cavity while the system is kept at EP. The results are obtained by analytical analysis and confirmed by 3D full wave simulations.

------

*Correspondence should be addressed to renminma@pku.edu.cn.

Optical vortices are light beams with an azimuthal phase dependence of $\exp(il\varphi)$. It possesses two important features: (1) It carries an quantized intrinsic orbital angular momentum (OAM) $l\hbar$, where $\hbar$ is the reduced Plank constant, $l$ is an arbitrary integer [1]; (2) It presents a topological phase singularity at the beam axis where the field amplitude vanishes due to the undetermined phase [2, 3]. Compared with the circular polarization beam that has only two states with spin angular momentum $\pm\hbar$, the OAM number of the vortices can take arbitrary integer number. Thus, it inspires wide attention to apply vortex beams to quantum information [4, 5] and communication [6, 7]. On the other hand, the special intensity null in the center of the beam made it useful in super-resolution imaging [8]. In addition, optical vortices have been explored to be used as optical tweezers [9] and optical spanners [10], laser manipulations [11], optical measurement [12, 13], digital imaging [14].

Conventional ways to generate optical vortex beam needs two components, a laser and a phase modulator. To date, various methods have been studied for wave-front phase modulation of laser emission to generate vortex beams, including spiral phase plates [15, 16], mode transformer [1], holograms [3, 17], inhomogeneous birefringent elements [18], and metasurfaces [19, 20], Pancharatnam–Berry phase optical elements [21], and devices to couple the conventional light into a micro-ring with angular gratings [22]. Recently, chromophore nanoarrays, chiral arrangement of molecular nanoemitters and topological defects in photonic crystals have been proposed to directly generate vortex beams [23, 24, 25, 26]. However, a vortex laser with well-defined vortex beam emission remains a fundamental challenge. In stark contrast with normal standing wave based lasers, a vortex laser needs a travelling lasing mode and precisely determined phase of the emission at each position on the cavity.

The exceptional point is a singular point of non-Hermitian system at which not only the eigenvalues but also the eigenstates coalesce [27]. Research in EP has been intensified as a result of its drastic physical behavior and application in novel devices such as unidirectional reflectionless light propagation [28-34], directional lasing [35], pump-induced turn off of the laser [36], loss induced lasing [37], mode interchange while encircling an EP [38], charity of the eigenmode [39-41], and single-particle detection [42, 43]. In this report, we proposed a vortex laser using exceptional point in a non-Hermitian system that can directly emit vortex beam with definite OAM. The refractive index modulation and along azimuthal direction combined with the grating protruding from the micro-ring cavity modulate the eigenmode of a micro-ring cavity to be a vortex laser mode.

Fig. 1(a) shows the schematic of the vortex beam laser. The micro-ring cavity has outer radius $R$, width $W$, and height $L$. The active region (micro-ring) is surrounded by air. Position in the system is specified by cylindrical coordinate $(\rho, \varphi, z)$, in which $\rho$ is the polar radius, $\varphi$ is the azimuthal angle, and $z$ is the distance from the ring plane. The plane $(z = 0)$ locates at the middle of the micro-ring. This coordinate system is illustrated in Fig. 1. The micro-ring cavity is azimuthally modulated in refractive index and periodically grated along the outer wall. For example, we chose InGaAsP as the gain material, for its emission lying at the C-band of the optical communication [44]. In addition, the index modulation along the azimuthal direction can be realized by evaporating additional materials on top of the cavity [45]. Artfully modulated refractive index tunes the system to EP, at which two degenerate counterpropogating cavity eigenmodes coalesce to a travelling whispering gallery mode (WGM), as shown in Fig. 1(b). The fluctuation of $|H|$ is very weak, which demonstrates a travelling wave mode. The grating elements protrude from the micro-ring cavity by $\Delta r = 60\ nm$, and their angular width is $\delta\varphi =0.03$ in radian. They equally space around the cavity. The travelling WGM is tuned to couple with a vertically emitting vortex beam with definite OAM by the grating elements. The OAM taken by the vortex beam can be tuned by the azimuthal order of the WGM and the number of grating elements around the cavity. Under uniform pumping, the travelling WGM loss (due to absorption and radiation) is compensated by the gain, and the lasing mode emits vortex beam into free space as shown by Fig. 1(c).

Here we explain how to modulate the system to EP. In order to show that we have many options to realize EP in the microring system, we adopted a general form of the refractive index along the azimuthal direction ($\varphi$) as (See Fig. 2(a)),

$$\Delta n = \begin{cases} n_0 + \Delta_1 e^{i\phi_1} & (l\pi/m \leq \varphi \leq l\pi/m + \delta\varphi_1) \\ n_0 + \Delta_2 e^{i\phi_2} & (l\pi/m + \varphi_0 \leq \varphi \leq l\pi/m + \varphi_0 + \delta\varphi_2) \end{cases} \quad (1)$$

where $n = 0,1,2,\ldots,2m-1$. The microring is divided into $2m$ periods. $n_0$ is the unperturbed part of the refractive index. The index modulation is given by complex number $\Delta_2 e^{i\phi_2}$ ($\Delta_1 e^{i\phi_1}$).

The index perturbation induces in the coupling between the two degenerate counterpropagating WGMs. Based on the coupled mode theory with a slowly varying approximation, for a negligible scattering loss of the micro-ring, the coupled mode equations of the desired WGM subvectorspace with azimuthal order of $m$ are

$$\begin{cases} \dfrac{dA}{Rd\varphi} = H_{11}A + H_{12}B \\ \dfrac{dB}{Rd\varphi} = H_{21}A + H_{22}B \end{cases} \quad (2)$$

where

$$H_{11} = -i2m\kappa(\Delta_1 e^{i\phi_1}\delta\varphi_1 + \Delta_2 e^{i\phi_2}\delta\varphi_2),$$

$$H_{12} = -\kappa[\Delta_1 e^{i\phi_1}(e^{i2m\delta\varphi_1} - 1) + \Delta_2 e^{i\phi_2}(e^{i2m(\varphi_0+\delta\varphi_2)} - e^{i2m\varphi_0})],$$

$$H_{21} = -\kappa[\Delta_1 e^{i\phi_1}(e^{-i2m\delta\varphi_1} - 1) + \Delta_2 e^{i\phi_2}(e^{-i2m(\varphi_0+\delta\varphi_2)} - e^{-i2m\varphi_0})],$$

$$H_{22} = i2m\kappa(\Delta_1 e^{i\phi_1}\delta\varphi_1 + \Delta_2 e^{i\phi_2}\delta\varphi_2),$$

$R$ is the resonator radius of the eigenmode, $\kappa = \dfrac{\omega^2 n_0 R}{2\pi c^2 m}$, $\omega$ is the degenerate resonance frequency without index perturbation modulation. The eigenmodes are given by $\psi = Ae^{im\varphi} + Be^{-im\varphi}$, where $A$ and $B$ are the amplitudes of the CW and CCW base modes, respectively.

$H_{12}$ ($H_{21}$) describes the coherent backscatterings of light from CCW mode to CW mode (from CW mode to CCW mode). When the backscatterings from CCW mode to CW mode under the index perturbation interfere destructively but the backscatterings from CW mode to CCW mode under the index perturbation don'tt interfere destructively, i.e., $H_{12} = 0$ and $H_{21} \neq 0$, the system locates at the Exceptional point (EP), at which not only the eigenvalues but also the eigenvectors coalesce and the remaining eigenstate is a pure CCW mode. In the same way, for $H_{21} = 0$ and $H_{12} \neq 0$, the system also locates at the EP and the remaining eigenstate is a pure CW mode. According to these conditions, we can obtain the parameters to realize EP.

Now, we show two classes of EP satisfying $H_{12} = 0$ and $H_{21} \neq 0$. If two modulation sections have the same angular width $\delta\varphi_1 = \delta\varphi_2$, the refractive index modulation for EP just needs to satisfy simple relations $\Delta_1 = \Delta_2$ and $\phi_2 - \phi_1 = \pi - 2m\varphi_0$. another class of refractive index modulation is the pure real part and imaginary part modulation in the two kinbds of sections respectively ($\phi_1 = 0$ and $\phi_2 = \pi/2$). Fig. 2 (b) and (c) show the relations the parameters need to satisfy. The systems with parameters locating at the surface in the graph are at EP. The PT-symmetrical refractive index modulation is the special condition shown in Fig. 2(b) and (c). In order to further demonstrate the results, we run a 2D simulation for the system of pure real part index modulation $n_R$ ($\phi_1 = 0$) and pure imaginary index modulation $n_I$ ($\phi_1 = \pi/2$) respectively, under the condition of $2m\varphi_0 = \pi/2$, as shown by Fig.2 (d) and (e). Obviously, the systems corresponding to the parameter surface over the clinodiagonal of the $n_R - n_I$ coordinate plate, which satisfy $\Delta_1 = \Delta_2$, have the coalesced eigenvalues and located at EP.

In the following, we adopt the PT-symmetricaly distribution of the refractive index to demonstrate the realize of the vortex beam laser. The PT-symmetrically refractive index along the azimuthal direction ($\varphi$) is:

$$n = \begin{cases} n_0 + \Delta n_R & (n\pi/m \leq \varphi \leq n\pi/m + \pi/4m) \\ n_0 + \Delta n_R + \Delta n_I i & (n\pi/m + \pi/4m \leq \varphi \leq n\pi/m + \pi/2m) \\ n_0 + \Delta n_I i & (n\pi/m + \pi/2m \leq \varphi \leq n\pi/m + 3\pi/4m) \end{cases} \quad (3)$$

where $n_0$ is the unperturbed part of the refractive index (the InGaAsP index

$n_0 = 3.42$), $\Delta n_R$ and $\Delta n_I$ are the real and imaginary index modulation, respectively. $n = 0,1,2,\ldots,2m-1$. The micro-ring is divided into $2m$ periods. In addition, we design the protrudent grating $q_{OWG} > m$ ($q_{OWG} \leq 25$). In this case, the protrudent grating does not cause the change of the EP in parameter space. It only takes the role of coupling the travelling WGM at EP into free space beam, which is the vortex beam with angular momentum index being solely determined by the difference between integers $m$ and the number of the protrudent gratings.

Firstly, we demonstrate the roles the refractive index modulation and grating along the ring cavity play. The refractive index modulation implements the EP while the protrudent grating leads to the coupling between the WGM and a vertically emitting vortex beam mode. A clockwise or counter-clockwise WGM traveling in the azimuthal direction varys as $\exp[i(\beta_{WGM}l_\varphi - \omega t)]$. This wave interacts with the azimuthal periodic grating or the modulated refractive index (period $\Lambda$) that can be represented as Fourier series [47]

$$\sum_{s=-\infty}^{\infty} A_s(r,z)\exp(i\,2\pi s l_\varphi/\Lambda) \qquad (4)$$

Thus, the traveling WGM is coupled to waves with

$$\beta_{s\varphi} = \beta_{WGM} + 2\pi s/\Lambda \qquad (5)$$

Among them, the waves satisfying the following condition can radiate into the free space,

$$|\beta_{s\varphi}| < 2\pi/\lambda \qquad (6)$$

where $\lambda$ is the wavelength in air. Then, it can be obtained that the traveling WGM is extracted into air region and forms the vortex beam when the following angular phase-matching condition is satisfied [22]

$$v_{rad} = m - gq \quad (7)$$

and

$$(n_{eff} - 1)\frac{2\pi R}{q\lambda} < g < (n_{eff} + 1)\frac{2\pi R}{q\lambda} \quad (8)$$

where $v_{rad}$ is the angular momentum index of the radiation beam, $|m|$ is the number of optical periods of the WGM around the resonator. $q$ represents the number of the protrudent grating elements ($q_{OWG} > |m|$) along the micro-ring cavity or the modulated refractive index cycles ($q_{IMP} = 2|m|$), $n_{eff}$ is the effective index of the WGM in the waveguide, $R$ is the resonator radius. Thus, we can design the cavity parameters to obtain an outgoing vortex beam with definite angular momentum index. For example, In our devices, $\lambda$ is around 1550 nm, $|m|$ is 16, $n_{eff}$ is around 2.6, $R$ is around 1750 nm. In this case, $|g|$ can only be 1. i.e.,

$$v_{rad} = m - q \quad (9)$$

Eq. (3) and (6) imply that the protrudent grating elements ($q_{OWG}$) takes the role of couplingthe WGM into free-space vortex beam mode with definite angular momentum index $v_{rad} = m - q_{OWG}$. Meanwhile, The index modulation ($q_{IMP} = 2|m|$) will not couple additional vortex beam into free space, due to it doesn't satisfy Eq. (3). This ensures that we can obtain a vortex beam with definite OAM.

On the other hand, the coupling of the CCW and CW WGM with angular momentum index $m$ occurs for [46]

$$2m = gq \quad g = \pm 1, \pm 2, \ldots \quad (10)$$

Eq. (7) shows that the index modulation ($q_{IMP} = 2|m|$) can induce the backscattering between the desired CCW and CW WGM with angular momentum index $m$. Thus, index modulation takes the role of tuning the system to EP. In addition, protrudent grating elements will not cause the change of the EP in parameter space since $2m \neq gq_{OWG}$. This ensures that we can tune the OAM of the emitting vortex beam by changing the number of the protrudent grating elements, making sure the system still at EP at the same time.

The general results obtained are suitable for both TE WGM and TM WGM. In our device with thin ring geometry, the effective index for TM modes is considerably decreased [48]. Thus, it is tuned that only TE modes preferentially reach the lasing condition. The TE modes has the magnetic field vector $H_z$ perpendicular to the cavity plane. The outer-wall near field of the TE WGM is scattered by the protrudent grating elements [22]. It has been demonstrated that the radiated beams are vector vortices with an OAM $(m - q_{OWG})\hbar$, where $\hbar$ is Planck constant $h$ divided by $2\pi$ [49].

Fig. 3 show that we can obtain the vortex beams with arbitrary OAM. The protrudent grating elements with $2m \neq gq_{OWG}$ do not cause the change of the EP in parameter space. Thus, with the increase of $q_{OWG}$, we can obtain the vortex beam with increasing OAM as shown by Fig. 3. Fig. 3 (a), (e), and (i) show the WGM intensity distribution $|H|$ in the cross section $z = 0$ for different grating elements ($q_{OWG} = 17, 18, 19$) at $\Delta n_R = \Delta n_I = 0.01$. The almost homogeneous field intensity distributions illustrate the perfect travelling mode properties. It means that changing the number of the grating elements hardly cause the change of the EP in the parameter space, since the grating doesn't induce the coupling between the CW and CCW modes (See Eq. (10)). Fig. 3 (b), (f), and (j) present the corresponding field intensity distribution $|H|$ in the cross section $z = 4105\ nm$. It can be seen that the fluctuation of $|H|$ is very weak. This demonstrates that the Vortex beam coupled with a traveling WGM has definite OAM $v\hbar$ and the component of the vortex beam with OAM $-v\hbar$ is negligibly small. The transversal distributions of radial component $H_r$ (Fig. 3 (c), (g) and (k)) and the corresponding phase distributions arg ($H_r$) (Fig. 3 (d), (h) and (l)) further confirm the vortex beam scattered by the protrudent grating elements with $q_{OWG} = 17, 18, 19$ has a definite OAM $\hbar$, $2\hbar$, and $3\hbar$, respectively, which can be calculated from Eq. (9).

The emitted vortex beam can have azimuthal, radial, and longitudinal filed components with exactly defined optical OAM. All components have an azimuthal phase dependence of $exp(il\varphi)$ identifying the OAM carried by the vortex beam [49]. For $|l| \neq 1$, the vortex beam have on-axis intensity null, resulting from the singularity where the local field is undefined (See Fig. 3 (f) and (j)). This intensity null has been explored to be applied to super-resolution imaging [8]. In contrast with scalar vortex beams, the cylindrical vector beam has a bright center when $|l| = 1$ [49], just like the simulation result [See Fig. 3 (b)].

Now, we illustrate the creation of the vortex beam by expanding wave function inside the ring cavity in cylindrical harmonics [48, 50],

$$H_z(\rho,\varphi) = \sum_{m=-\infty}^{\infty} \alpha_m J_m(nkr)exp(im\varphi) \quad (14)$$

Where $J_m$ is the $m$th-order Bessel function of the first kind, $k$ is the wave number, $n$ is the effective refractive index of the micro-ring. The CCW (CW) traveling-wave components is denoted by positive (negative) values of the angular momentum index $m$ [50]. For the system with $\Delta n_I = 0.01$, the ratio of the clockwise (CW) and counterclockwise (CCW) traveling waves component $|\alpha_m|^2/|\alpha_{-m}|^2$ as a function of real index modulation $\Delta n_R$ is shown in Fig. 4(c). At $\Delta n_R = 0$, both eigenmodes have equal CW and CCW components ($|\alpha_{16}|^2/|\alpha_{-16}|^2 \sim 1$). They are not the traveling wave modes but the standing wave modes. In the vicinity of the EP ($\Delta n_R = \Delta n_I$), both eigenmodes have dominant CCW component ($|\alpha_{16}|^2/|\alpha_{-16}|^2 \gg 1$), i.e., both eigenmodes have the same chirality and propagates in the same CCW direction. Especially, the simulation shows that the CCW component is about 484 times larger than the CW component at EP ($\Delta n_R = \Delta n_I$), which indicates a perfect traveling wave mode. The inset of Fig. 4(a) shows the simulated intensity patterns $|H|$ of the pair of modes in the cross section $z = 0$ at $\Delta n_R = \Delta n_I$. It can be seen that the fluctuation of $|H|$ is extremely weak. This further confirms that the mode in cavity is a travelling wave mode.

Fig. 4(d) presents the angular momentum distribution $|\alpha_m|^2$ of the WGM at EP ($\Delta n_R = \Delta n_I = 0.01$). It can be seen that the angular momentum distribution of the mode is dominated by the CW component ($m = 16$). Based on the Eq. (9), the protrudentprotrudent grating elements ($q_{OWG} = 18$) extracts CW and CCW traveling modes to the vertically emitted vortex beams with OAM index $m = -2$ ($16 - 18$) and $m = 2$ ($-16 + 18$), respectively. The CW component is two orders larger than the CCW component ($m = 16$). Thus, the emitted vortex beam with $m = -2$ from CW mode is two orders larger than the vortex beam with $m = 2$ from CCW mode, just as shown by Fig. 4(d). Obviously, a vortex beam with definite angular momentum is obtained.

At last, we illustrate the vortex beam is robust in the lasing process under pumping. The uniform pumping of the gain material (InGaAsP) of the cavity produces a uniform increase of the imaginary part of refractive index of the InGaAsP. It results in the uniformly increasing of the imaginary part of background refractive index $n_I$. For Eq. (3), a change of the background refractive index $n_I$ will not cause the change of the index Fourier expansion coefficient with grating vector $2\pi s/\Lambda \neq 0$. Based on the angular phase matching condition, it doesn't cause the coupling between the CCW and CW WGM with OAM $m\hbar$ and $-m\hbar$ respectively, i.e., the uniform pumping doesn't cause the change of the EP in parameter space.

Fig. 5 shows the 3D full wave simulation results of the gain effect in a vortex beam laser device at EP. The uniform pumping of gain material (InGaAsP) produced gain increasing process of the InGaAsP ring. It is mimicked by increasing the imaginary part of background refractive index $n_I$. The quality factor is about 365 for the cavity without gain. With the increasing of the gain coefficiency, the cavity quality factor increases by orders of magnitude, indicating that the loss is compensated by the gain (See Fig. 5 (a)). Then, the mode becomes a lasing mode and emits vortex beam. The corresponding angular momentum distribution of the mode $H_z$ in the cross section $z = 0$ is shown in Fig. 5 (b). The black filled circles show the main component of the WGM is CW. The CW traveling mode is scattered by the grating outside of the InGaAsP ring, which leads to a vortex beam with azimuthal quantum number $m = -2$ (see the red filled circles). Their OAM components are almost unchanged with the increase of the background refractive index $n_I$. This further confirms that the vortex beam laser is stable in the lasing process, as predicted in the aforementioned theoretical analysis.

In summary, we proposed to use exceptional points to realize a micro vortex laser. It can directly generate optical vortex radiation with well-defined OAM at will. We believe that our approach not only boosts large-scale integrated applications related to vortex beam but also motivates the research on the physics and application of exceptional points in optical field.

*The authors' note at submission: after the completion of this work, we found a similar work on Science just published online today (Science, 353, 464-467, 2016).*

**Acknowledgements**

This work was supported by the "Youth 1000 Talent Plan" Fund, Ministry of Education of China (No. 201421) and the National Natural Science Foundation of China (Nos. 11574012, 61521004).

momentum states of photons. Nature 412, 313–316 (2001).

[5] G. Molina-Terriza, J. P. Torres, L. Torner, Nat. Phys. 3, 305 (2007).

[6] Jian Wang, Jeng-YuanYang, Irfan M. Fazal, Nisar Ahmed, Yan Yan, HaoHuang, Yongxiong Ren, Yang Yue, Samuel Dolinar, MosheTur and Alan E. Willner, Terabit free-space data transmission employing orbital angular momentum multiplexing, NATURE PHOTONICS, 6, 488 (2012).

[7] Xiannan Hui, Shilie Zheng, Yiling Chen, Yiping Hu, Xiaofeng Jin, Hao Chi, and Xianmin Zhang, Multiplexed millimeter wave communication with dual orbital angular momentum (OAM) mode antennas, Scientific Reports, 5, 10148, (2015).

[8] F. Tamburini, G. Anzolin, G. Umbriaco, A. Bianchini, and C. Barbieri, "Overcoming the Rayleigh Criterion Limit with Optical Vortices," Phys. Rev. Lett. 97(16), 163903 (2006).

[9] H. He, N. R. Heckenberg, and H. Rubinsztein-Dunlop, J. Mod. Opt. 42, 217 (1995).

[10] M. E. J. Friese, J. Enger, H. Rubinsztein-Dunlop, and N. R. Heckenberg, Phys.Rev.A 54, 1593 (1996).

[11] K. Toyoda, K. Miyamoto, N. Aoki, R. Morita, and T. Omatsu, Nano Lett. 12, 3645 (2012).

[12] Z. Y. Zhou, Y. Li, D. S. Ding, W. Zhang, S. Shi, and B. S. Shi, "Optical vortex beam based optical fan for high-precision optical measurements and optical switching," Opt. Lett. 39(17), 5098–5101 (2014).

[13] Fürhapter, S., Jesacher, A., Bernet, S. & Ritsch-Marte, M. Spiral interferometry. Opt. Lett. 30, 1953–1955 (2005).

[14] Torner, L., Torres, J. & Carrasco, S. Digital spiral imaging. Opt. Express 13, 873–

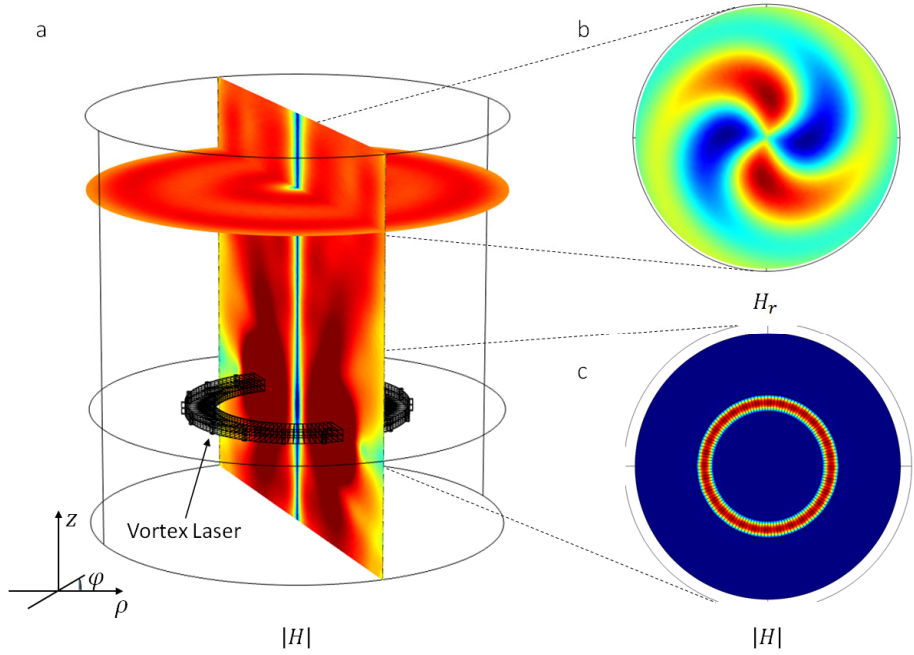

**Figure 1** The schematic of the vortex laser. (a) the perspective view and operation principle of the vortex laser with the distribution of the magnetic field $|H|$ on log scale. (b) The intensity distribution of the magnetic field $|H|$ in the cross section of the ring cavity. (c) The transversal distributions of radial component of the magnetic field $H_r$ in the cross section of the vortex beam.

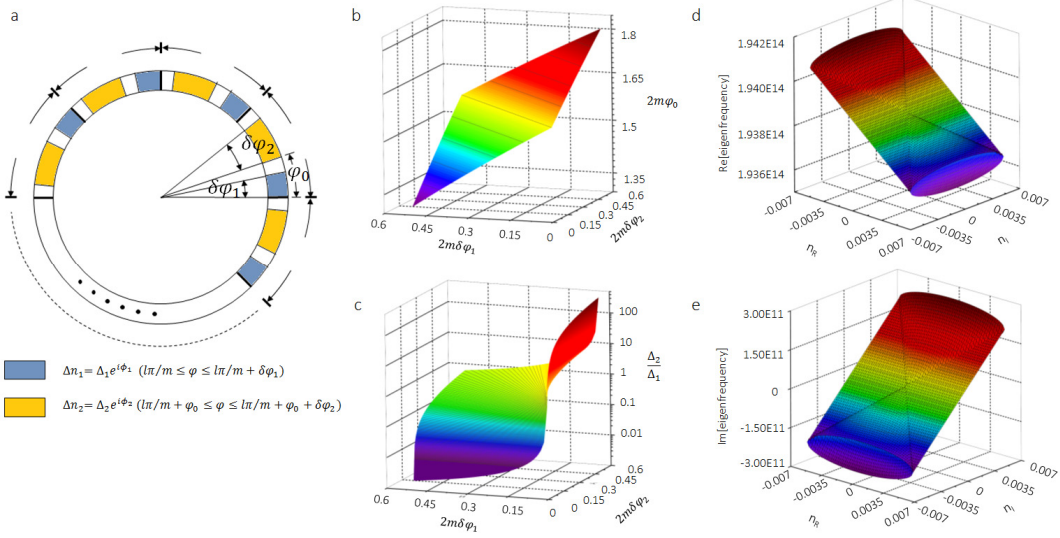

**Figure 2** (a) The schematic of the refractive-index modulation of a general form. (b) and (c) The two conditions of the parameters to be meeted simultaneously to achive EP. The real part (d) and imaginary part (e) of the eigenfrequencies with azimuthal order $m = 16$ in a 2D ring cavity as a function of the refractive index modulation $\Delta n_R$ and $\Delta n_I$. Here, the refractive index of the cavity is $n_0 = 2.67$ with a modulation of the real part $\Delta n_R$ in $n\pi/16 \leq \varphi \leq n\pi/16 + \pi/32$ and imaginary part $\Delta n_I$ in $n\pi/16 + \pi/64 \leq \varphi \leq n\pi/16 + 3\pi/64$, respectively. $n = 0,1,2,...,31$. The refractive index out of the cavity is 1. The width of the cavity is 500 nm. The inner radius of the cavity is 1500 nm.

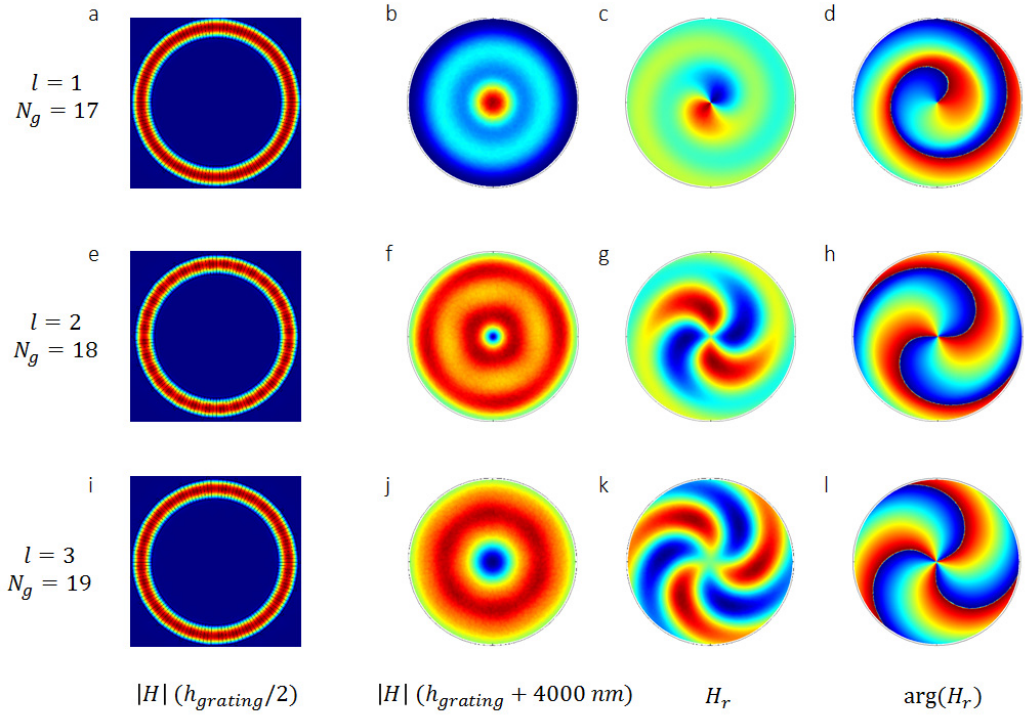

|H| (h_{grating}/2)   |H| (h_{grating} + 4000 nm)   $H_r$   $\arg(H_r)$

**Figure 3** Calculated transversal characteristics of the modes with azimuthal order $m = 16$ in the systems with $N_g = 17$ (first row), $N_g = 18$ (second row) and $N_g = 19$ (third row). All systems have the same index modulation ($\Delta n_R = \Delta n_I = 0.01$). The first column presents the magnetic field intensity distributions ($|H|$) in the cross section at the middle of the InGaAsP microring ($z = 0$). The magnetic field intensity distributions ($|H|$), the radial component of the magnetic field ($H_r$) and the phase of the radial component of the magnetic field ($\arg(H_r)$) in the cross section at far field ($z = 4105\ nm$) are presented in the second, the third and the fourth column, respectively.

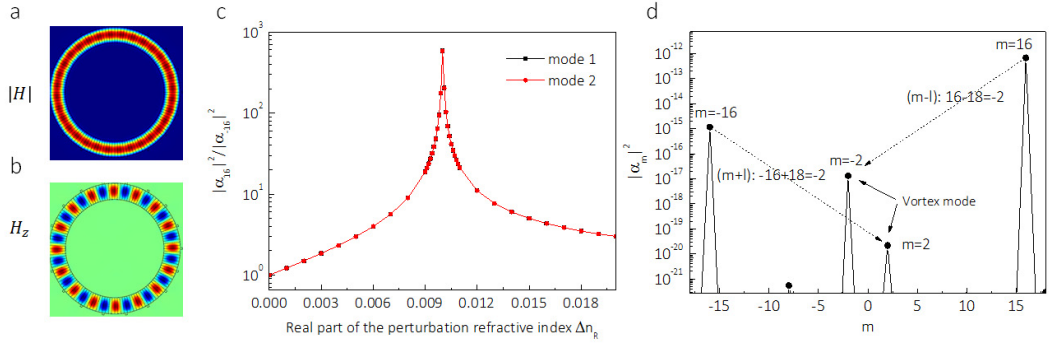

**Figure 4** Angular momentum distributions $|\alpha_m|^2$ as a function of real index modulation $\Delta n_R$. (a) and (b) show the simulated intensity patterns $|H|$ and $H_z$ of a mode pair in the cross section $z=0$ at $\Delta n_R = \Delta n_I$. (c) The imaginary index modulation is fixed at $\Delta n_I = 0.01$. (a) Ratio of the clockwise (CW) and counterclockwise (CCW) traveling waves component $|\alpha_m|^2/|\alpha_{-m}|^2$ as a function of real index modulation $\Delta n_R$. (d) Angular momentum distribution $|\alpha_m|^2$ of the WGM at EP ($\Delta n_R = \Delta n_I = 0.01$).

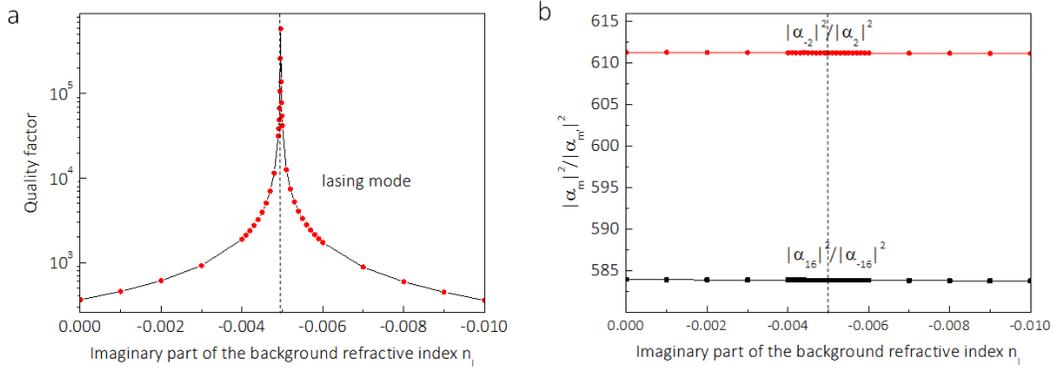

**Figure 5** Loss compensation in a vortex beam laser at EP ($\Delta n_R = \Delta n_I = 0.01$) with $N_g = 19$. (a) the background gain dependence of the cavity quality factor. The uniform pumping produced gain increasing process of the InGaAsP ring is mimicked by increasing the imaginary part of background refractive index $n_I$. The quality factor is about 365 for the cavity without gain. With the increasing of the gain coefficient, the cavity quality factor increases by orders of magnitude, indicating that the loss is compensated by the gain. (b) Ratio of the clockwise (CW) and counterclockwise (CCW) traveling waves component $|\alpha_m|^2/|\alpha_{-m}|^2$ as a function of background refractive index $n_I$. The black filled circles show the main component of the mode is CW mode, which is almost unchanged with the increase of the background refractive index $n_I$. The CW traveling mode is coupled to a vortex beam with azimuthal quantum number $m = -2$ (see the red filled circles).